\documentstyle[pre,aps,epsf]{revtex}
\begin{document}
\draft
\title{Multiparticle trapping problem in the half-line}
\author{Santos B. Yuste\thanks{Permanent address: Dpto. de F\'{\i}sica,
Univ.\ de Extremadura, E-06071 Badajoz, Spain}}
\address{Department of Chemistry 0340, University of
California San Diego, La Jolla, CA 92093, USA}
\author{L. Acedo} \address{Departamento de F\'{\i}sica,
Universidad  de Extremadura, E-06071 Badajoz, Spain}
\date{January 25, 2001}
\maketitle
\begin{abstract}
A variation of Rosenstock's trapping model in which $N$
independent random walkers are all initially placed upon a site of a
one-dimensional lattice in the presence of a {\em one-sided} random
distribution (with probability $c$) of absorbing traps is investigated.
The  probability (survival probability)  $\Phi_N(t)$ that no random walker is trapped by time $t$
for $N \gg 1$ is calculated by using the extended Rosenstock approximation.
This requires the evaluation of
the moments of the number $S_N(t)$ of distinct sites
visited in a {\em given} direction up to time $t$ by $N$ independent random
walkers.
The Rosenstock approximation improves when $N$ increases, working well in the range
$Dt\ln^2(1-c) \ll \ln N$, $D$ being the diffusion constant.  The moments  of the time (lifetime)
before any trapping event occurs are calculated asymptotically, too. The agreement with numerical
results is excellent.

\end{abstract}
\pacs{PACS: 05.40.-a, 66.30.-h, 02.50.Ey\\
Keywords: Rosenstock trapping model; Rosenstock approximation; Multiparticle diffusion problems; Survival probability
}

\section{Introduction}
\label{sect_1}
Survival of Brownian particles in a medium populated with randomly distributed
static traps is a fundamental problem (the ``trapping'' problem) of random walk theory 
that has been an active area of research for decades with many applications
in physics and chemistry \cite{Hughes,Weiss,HollanderWeiss,ShDba}. 
The origin of this problem can be traced back to Smoluchowski's theory
of coagulation of colloidal particles \cite{Hughes,Weiss,HollanderWeiss}. It has now
become a basic model of widespread interest in areas such as trapping of mobile defects
in crystals with point sinks \cite{BeelerDamask,Rosens}, the kinetics of luminescent
organic materials \cite{Rosens}, the kinetics of photosynthetic light energy
to oxygen conversion \cite{Montroll}, anchoring of polymers by chemically
active sites \cite{Oshanin}, atomic diffusion in glasslike materials
\cite{Miyagawa} and many more \cite{Klafter}.

This paper is devoted to a variation of the so-called Rosenstock trapping
problem on a one-dimensional  substrate. Usually, the
one-dimensional Rosenstock trapping problem (which we will call the ``two-sided''
Rosenstock trapping problem for reasons that will be apparent later on) is
stated as follows \cite{Hughes,Weiss,HollanderWeiss}. A one-dimensional lattice
is filled with a random distribution of static traps; then,  one ($N=1$) random
walker is placed initially ($t=0$) at a given site of the lattice; it starts to
diffuse and, eventually, is caught by a trap.
In this paper  we study a different but closely related trapping problem (which we will call the ``one-sided'' Rosenstock
trapping problem) in which  (i) only a {\em half-line} of a one-dimensional lattice
is filled with a random distribution of static traps with concentration $c$ (this process could mimic
the excitation or production of defects in one side of a fiber by irradiation,
the other side being shielded) and (ii)  $N$ independent random
walkers are placed initially ($t=0$) at the contact point ($x=0$) between the two
half-lines that is taken as origin. These random walkers start to diffuse,
and eventually one of them is trapped at the nearest site occupied by a trap
(or deactivates it).  The statistical quantities of main interest are
the survival probability, $\Phi_N(t)$, defined as the probability that no random walker
has been trapped by time $t$, and the lifetime, $T_N$, defined as
the average time at which the first random walker of the set of $N$ arrives at
a trap site.
To our knowledge, this is the first multiparticle ($N\neq 1$) Rosenstock trapping problem ever studied.
A good reason for this is that the multiparticle versions of the trapping  problem  are much more difficult to solve
than  the trapping problems with a single particle.

This fact will be evident is this paper: the present trapping problem is elementary for  $N=1$
 and, in this case,  we will report the main results for the sake of completeness only (see section \ref{singleRW}).
 On the other hand, the multiparticle version is much more involved (see section \ref{onesidedN} and \ref{NRW}).
As an  exact evaluation of $\Phi_N(t)$ for $N > 1$ is  elusive,
 we have resorted to asymptotic analysis techniques. In particular, for $N
\gg 1$, we have used the extended Rosenstock approximation (or truncated
cumulant approximation) \cite{Hughes,Weiss,Klafter,Zumofen,Blumen}. This
requires to find the first moments $\langle S^m_N(t) \rangle$, $m=1,2,\ldots$, of $S_N(t)$, the number of sites situated to
the `right' of the origin $x=0$ that were visited up to time $t$ by $N$ random
walkers which started at the site $x=0$ at time $t=0$. Note that $S_N(t)$ is
just the maximum distance reached by any of the $N$ random walkers in the $+x$
direction from the origin by time $t$, i.e., $S_N(t)$ is the one-sided span of
the $N$-particle random walk.
The problem of evaluating  the first moment
$\langle S_N(t) \rangle$ has already been addressed in \cite{SA,KR,KR-Am} explicitly and
 in  \cite{LarraldeNydemas,LarraldeP,PRE-RCEuFrac}  implicitly.
The quantity  studied in these last references  was the number of distinct sites explored by $N$
random walkers on the one-dimensional lattice in {\em either} direction, $\widetilde S_{N}(t)$, but
$\langle S_N(t)\rangle$=$\langle \widetilde S_{N}(t)\rangle/2$.
However, little is known about higher moments of $S_N(t)$,  $\langle S^m_N(t)\rangle$,
except for some rough estimates \cite{KR,KR-Am,LarraldeP}.
The idea of evaluating the survival probability for the multiparticle
trapping problem by using the  moments of $\widetilde S_{N}(t)$  into the Rosentock approximation was suggested by
Larralde et al.\ in  \cite{LarraldeP}, although, to the best of our knowledge, it has
not been implemented perhaps for the lack of precise expressions for the
moments  $\langle \widetilde S^m_{N}(t)\rangle$.
Therefore, in order  to use the Rosenstock approximation in our one-sided multiparticle trapping problem,
we must  find  rigorous asymptotic results for the moments of $S_N(t)$. This is another objective (important in itself)
of this present work that, besides, we hope can illuminate
 how to deal with the evaluation of the moments of $\widetilde S_N(t)$  for other substrates.

The paper is organized as follows.
In section \ref{OneSidedRosensock} we recall the connection between the survival probability
$\Phi_N(t)$ and the so-called fixed trap survival probability (i.e., the
survival probability when the trap is placed at a given distance) and give the
basics of the extended Rosenstock approximation. The one-sided trapping problem for $N=1$ is
addressed in section \ref{singleRW}. In section \ref{onesidedN} we calculate the moments of  $S_N(t)$
in the form of an asymptotic series in which the corrective terms decay as powers of $1/\ln N$.
The evaluation of $\langle S^m_N(t) \rangle$ is a necessary prerequisite for the
implementation, in section \ref{NRW}, of the Rosenstock approximation for  $\Phi_N(t)$.
In this section  we compare $\Phi_N(t)$ as given by
the extended Rosenstock approximation with numerical results.
The moments of the  lifetime $T_N$   and its variance
are evaluated  in section \ref{sec_lifetime}. The paper ends with
some conclusions and remarks.

\section{One-sided trapping model and Rosenstock approximation}
\label{OneSidedRosensock}

The one-sided Rosenstock trapping model is defined as follows:
(i) Quenched traps are randomly distributed on the right-hand side of a
one-dimensional lattice ($x > 0$) with concentration $c$ ($1-c\equiv p$);
(ii) the random walkers are placed initially upon site $x=0$ which divides
the randomly filled trapping half-line and the empty one; and
(iii) the traps are irreversible, that is, a walker encountering a trap
is killed there.
Then,  the survival probability is given by
$
\label{PhiNSum}
\Phi_N(t)=\sum_{r=1}^t  p^r P_N(t \vert r)  ,
$
where $P_N(t \vert r)$ is the probability that the span of the $N$ random
walkers in the positive direction (the largest distance reached by any
of the $N$ random walkers for $x > 0$) is equal to $r$ after $t$ time
steps \cite{Hughes}.
Let $\Gamma_N(t \vert r)$ be the probability that the site $x=r$
has not been visited by any of the $N$ random walkers by time $t$ (the
so-called fixed-trap survival probability).
Then, in the continuous limit,
\begin{equation}
\label{PhiNInt}
\Phi_N(t)=\int_0^\infty p^r \frac{d \Gamma_N(t \vert r)}{d r} d r\; ,
\end{equation}
where we have used the relationship $P_N(t \vert r)=d \Gamma_N(t \vert r)/ d r$  between the one-sided span
distribution $P_N(t \vert r)$ and the fixed-trap survival probability $\Gamma_N(t \vert r)$.
For $N$ independent random walkers one has
$\Gamma_N(t|r)=[\Gamma(t|r)]^N$  where $\Gamma(t|r)=\Gamma_1(t|r)$ is the probability that distance $r$ has not
been reached by a single random walker by time $t$.
For the one-sided diffusion process it is well-known that \cite{Hughes,Weiss}:
\begin{equation}
\label{GtrN1}
\Gamma(t \vert r)=\text{erf} \left( \frac{r}{\sqrt{4 D t}} \right) .
\end{equation}

The extended Rosenstock approximation (or truncated cumulant expansion) is now
a standard approach \cite{Hughes,Weiss,HollanderWeiss} to the Rosenstock
trapping problem, which was first proposed by Zumofen and Blumen \cite{Zumofen} and that we recall here for the sake of reference.
From the definition of $S_N(t)$ as the number of distinct sites on the positive half-line visited up to time $t$ by $N$
independent random walkers that started at $x=0$ at time $t=0$ (note that this means $S_N(0)=0$),
the survival probability of the $N$ random walkers  is
$
\Phi_N(t)=\left\langle p^{S_N(t)} \right\rangle \equiv \left\langle \exp[{S_N(t) \ln p]}
\right\rangle .
$
The average in this equation is performed
over all realizations of the random walkers' exploration of the lattice up to time step $t$.
The well-known cumulant expansion technique \cite{Hughes,Weiss}
allows an alternative form of $\Phi_N(t)$ as an infinite series
expansion \begin{equation}
\label{Ros:gen}
\Phi_N(t)=\exp \left[\sum_{n=1}^\infty \displaystyle\frac{\kappa_n (\ln p)^n}{n !}\right]
\end{equation}
where $\kappa_n$, $n=1,2,\ldots$ denote the cumulants of $S_N(t)$:
$\kappa_1 = \left\langle S_N(t) \right\rangle $,
$\kappa_2=\left\langle S_N^2(t) \right\rangle-\left\langle S_N(t) \right\rangle^2\equiv \sigma_N^2(t) $, \ldots
If only the first term of the sum in  Eq.\  (\ref{Ros:gen}) is kept we
arrive at the zeroth-order Rosenstock approximation
\begin{equation}
\label{Ros:zero}
\Phi_N^{(0)}(t)=e^{\left\langle S_N(t) \right\rangle \ln p }.
\end{equation}
The error made by using this approximation can be
estimated by taking into consideration the next exponential term in Eq.\
(\ref{Ros:gen}):
$\Phi_N(t)=\Phi_N^{(0)}(t)\left[1+{\cal O}\left(\sigma_N^2 \left( \ln p \right)^2
\right) \right] .
$
Thus, the condition $\sigma_N^2 \ll 1/(\ln p)^2$ must be fulfilled for the zeroth-order Rosenstock approximation to be reasonable.
The first-order (extended) Rosenstock approximation
is obtained by retaining two terms in the infinite sum of Eq.\
(\ref{Ros:gen}):
\begin{equation}
\label{Ros:first}
\Phi_N^{(1)}(t)=\exp\left[\displaystyle \left \langle S_N(t) \right \rangle
\displaystyle\ln p+\displaystyle\frac{1}{2}
\sigma_N^2 (\ln p)^2 \right].
\end{equation}
The relative error of this expression  is  of order ${\cal O}\left[\kappa_3 (\ln p)^3\right]$.

\section{One-sided trapping problem with a single random walker}
\label{singleRW}

The one-sided trapping problem is quite simple for $N=1$ .
Anyway, we report here the main results for the sake of completeness.
 From Eqs.\  (\ref{PhiNInt}) and (\ref{GtrN1})  the survival probability of a single random walker is
\begin{equation}
\label{PhiN1}
\Phi_1(t)=\int_0^{\infty} d r p^r \frac{d}{d r} \text{erf} \left( 
\frac{r}{\sqrt{4 D t}} \right)=e^{x^2/4} \text{erfc} \left( \frac{x}{2} \right)
\end{equation}
where $x=\sqrt{4 D t} \ln (1/p)$.
For very long times, $x \rightarrow \infty$, the asymptotic expansion of the
complementary error function \cite{Abramo} allows us to write
\begin{equation}
\label{PhiN1as}
 \Phi_1(t)=\frac{2}{\sqrt{\pi}} \frac{1}{x}
\left\{1+\sum_{m=1}^{\infty} (-1)^m 2^m \frac{(2 m - 1) !
!}{x^{2m}}\right\}
\end{equation}
where  $(2 m - 1)!!=(2m-1)\cdots 5\cdot 3 \cdot 1$.
Thus,  an asymptotic time regime is reached for
$t \gg 1/(\ln p)^2$
where the survival probability exhibits a power law decay 
$\Phi_1(t) \approx 1/[\sqrt{\pi D} \ln (1/p)] t^{-1/2}$.
This is an {\em algebraic} fluctuation slowdown corresponding to the
Donsker-Varadhan limit.

In order to apply the extended Rosenstock approximation to the single random walker case
for small $x$, we must evaluate the moments of the one-sided span $S_1(t)$:
\begin{equation}
\label{momenS+1}
\langle S^m_1(t) \rangle =
\int_0^\infty \frac{d\Gamma(t|r)}{dr}r^m dr
\end{equation}
or,  after integrating by parts,
\begin{equation}
\label{S1momen}
\left\langle S_1^m(t) \right\rangle=m\int_0^\infty \text{erfc}\left(
\frac{r}{\sqrt{4 D t}}\right) r^{m-1} d r=
\frac{\Gamma\left(\displaystyle\frac{m+1}{2}\right)}{\sqrt{\pi}} \left(4 D
t\right)^{m/2} \end{equation}
with $\Gamma(m)$ being the gamma or factorial function and where Eq.\ (\ref{GtrN1}) has been used.
It is now easy to verify that
$\kappa_n=a_n (4 D t)^{n/2}$, $n=1,2,\ldots$ with $a_1=1/\sqrt{\pi}$, $a_2=1/2-1/\pi$,
$a_3=(4-\pi)/(2 \pi \sqrt{\pi})$, $a_4=(2\pi-6)/\pi^2$, \ldots .
Direct substitution of these cumulants into the general expression for the Rosenstock
approximation (\ref{Ros:gen}) yields
\begin{eqnarray}
\label{Phi1:Ros}
\Phi_1(t)&=&\exp\left\{ \sum_{n=1}^\infty  \frac{a_n}{n!} \left(4 D 
t\right)^{n/2} \ln^n p
\right\}=\exp\left\{ \sum_{n=1}^\infty \frac{(-1)^n}{n!} a_n x^n 
\right\} \\
\noalign{\smallskip}
&=&\exp\left\{-\frac{1}{\sqrt{\pi}} x
+\frac{\pi-2}{4 \pi} x^2+\frac{\pi-4}{12 \pi \sqrt{\pi}} x^3+\frac{\pi-3}{12 
\pi^2} x^4 +{\cal O}(x^5)\right\} \nonumber
\end{eqnarray}
Notice that the Rosenstock approximation given by Eq.\  (\ref{Phi1:Ros}) coincides
with the exact result in Eq.\  (\ref{PhiN1}).

\section{One-sided span of a set of random walkers}
\label{onesidedN}

The objective of this section is twofold: first we want to obtain rigorous asymptotic expansions for
the one-sided span moments
\begin{equation}
\label{momenS+}
 \langle S^m_N(t) \rangle =
\int_0^\infty \frac{d\Gamma_N(t|r)}{dr}r^m dr
\end{equation}
for $m=1,2,\ldots$ and $N \gg 1$ independent random walkers, and, second, we want to check the reliability
of the obtained asymptotic expressions comparing them with numerical results.

Integrating Eq.\ (\ref{momenS+}) by parts, we find
\begin{equation}
\label{momenSxi}
\langle S^m_N(t)\rangle = m (4Dt)^{m/2}
 \int_0^\infty \xi^{m-1} \left[1-\text{erf}^N(\xi) \right] d\xi  .
\end{equation}
where $\xi=r/\sqrt{4Dt}$.  In order
to evaluate this integral for large values of $N$ it suffices to know $\Gamma(t|r)=\text{erf}(\xi)$ for large $\xi$, namely,
$\text{erf}(\xi)\approx
1-\pi^{-1/2} \xi^{-1} e^{-\xi^2}
\left(1-\xi^{-2}/2+\cdots\right)
$.
The asymptotic evaluation for large $N$ of the
 integral of Eq.\  (\ref{momenSxi}) is not a easy task. Fortunately, if one compares this integral with the
one carried out in Ref.\  \cite{PRE-RCEuFrac}, one realizes that both integrals are formally  equivalent.
In this way one  finds
\begin{equation} \langle S^m_N \rangle \approx
\left[4Dt \ln(N) \right]^{m/2} [1-\Delta(m)]
\label{Snt}
\end{equation}
where 
\begin{equation}
\Delta(m)=\sum_{n=1}^\infty \ln^{-n}N\sum_{j=0}^n 
s_j^{(n)}(m) \ln^j\ln N .
\label{Delta}
\end{equation}
Up to second order ($n=2$) the coefficients $s_j^{(n)}$ are 
\begin{eqnarray}
\label{s10}
s_0^{(1)}(m)&=& -\frac{m \omega}{2}  , \\
 s_1^{(1)}(m)&=& \frac{m}{4} , \\
s_0^{(2)}(m)&=&
\frac{m}{8}\left[(2-m)\left( \frac{\pi^2}{6}+\omega^2 \right)
+2\omega+2\right] ,\\
s_1^{(2)}(m)&=&  \frac{m}{8} \left[(m-2)\omega-1\right] ,\\
s_2^{(2)}(m)&=& \frac{m}{32} (2-m)  ,
\label{stilde}
\end{eqnarray}
where $\omega=\gamma-\case{1}{2}\ln\pi=0.0048507\cdots$ and 
$\gamma=0.577215\cdots$ is Euler's constant. 
In particular, the average value of $S_N(t)$ is given by  
\begin{equation}
\langle S_N \rangle \approx
\left[4Dt \ln(N) \right]^{1/2} \left[1-\frac{\ln\ln N-2\omega}{4\ln N}-
\frac{\case{1}{4}\ln^2\ln N-(1+\omega)\ln\ln N+\tilde s_0^{(2)}(1)}{8\ln^2 N}+\cdots \right]
\label{Snt1}
\end{equation}
with $\tilde s_0^{(2)}(1)=\pi^2/6+\omega^2+2\omega+2=3.654659\cdots$.
It is remarkable how close the result obtained in \cite{KR-Am}, namely,
\begin{equation}
\langle S_N \rangle \approx
\left[4Dt \ln(N) \right]^{1/2} \left[1-\frac{\ln\ln N+\ln(4\pi)}{4\ln N} +\cdots\right] ,
\label{Snt1KR}
\end{equation}
is to the rigorous result given by Eq.\  (\ref{Snt1}).
Notice that the result of Eq.\  (\ref{Snt1KR}) starts to differ from that of Eq.\   (\ref{Snt1})
in the first corrective term:
in this equation  $\ln (4 \pi)=2.531024\cdots$ plays the role of $-2\omega=-0.009701\cdots$ in Eq.\  (\ref{Snt1}).

\begin{figure}
\begin{center}
\leavevmode
\epsfxsize = 6.5cm
\epsffile{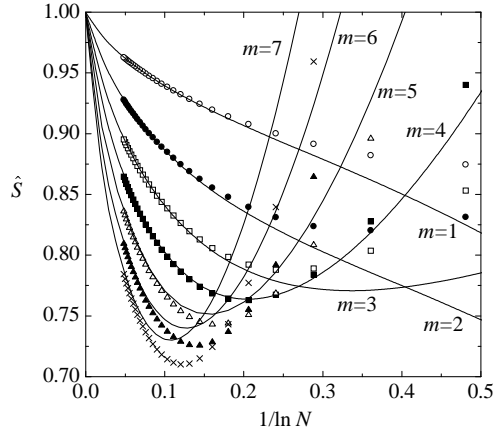}
\end{center}
\caption{
The dependence on $N$ of the first seven moments of $S_N(t)$.  We have plotted
${\widehat S}\equiv \langle S^m_N\rangle/[4Dt \ln N]^{m/2}$ versus $1/\ln N$.
The symbols correspond to the numerical estimate for $N=2^n$ with $n=3,5,\ldots
30$, when  $m=1$ (open circles), $m=2$ (filled circles), $m=3$ (open squares),
$m=4$ (filled squares), $m=5$ (open triangles), $m=6$ (filled triangles) and
$m=7$ (crosses).  The solid lines correspond to the second-order asymptotic
approximation $1-\Delta(m)$ as given by Eq.\   (\protect{\ref{Delta}}).
Notice the importance of the corrective terms as the main term (zeroth-order approximation)
 would be an horizontal line passing by ${\widehat S}=1$ }
\label{fig:momentos}
 \end{figure}
 \begin{figure}
\begin{center}
\leavevmode
\epsfxsize = 6.5cm
\epsffile{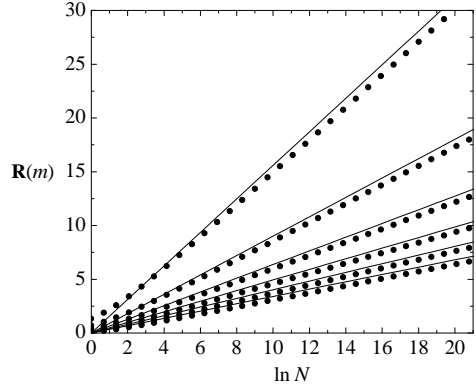}
\end{center}
\caption{
The ratio
$ R(m) \equiv\left[\langle S_N\rangle / \sigma_{N}(m)\right]^{m/2}$  versus $\ln N$ for
 (from top to bottom)  $m=2,3,\ldots,7$.
The symbols correspond to the numerical evaluation for $N=2^n$ with
$n=0,1,\ldots 30$.  The lines correspond to the asymptotic result of Eq.\   (\protect{\ref{pseudoratio}}).
The slope of the linear fit to the last five points ($N$ from
$2^{26}$ to $2^{30}$) is $1.53$ (1.56) for $m=2$, $0.88$ (0.90) for $m=3$, $0.62$ (0.64) for
 $m=4$, $0.48$ (0.49) for $m=5$, $0.39$ (0.40) for $m=6$, and $0.33$ (0.34) for $m=7$, in good agreement with the corresponding
asymptotic value $\{48/[m(m-1)\pi^2]\}^{1/2}$ given by Eq.\   (\protect{\ref{pseudoratio}}) which we have set in parentheses.
}
\label{fig:varratio}
\end{figure}

In Fig.\  \ref{fig:momentos} we compare the values of
$\langle S^m_N \rangle/[4Dt \ln N]^{m/2} $ for $m=1,2,\ldots 7$
 obtained by integrating Eq.\   (\ref{momenSxi}) numerically
with those obtained by means of the second-order asymptotic approximation of Eq.\   (\ref{Snt}).
The importance of the corrective terms given by $\Delta$ is evident as well as the good performance
(even for not-too-large values of $N$) of the second-order asymptotic expression.
This is especially notable for low-order moments. Notice that, at least for $n=1,2$,
the coefficient  $s_j^{(n)}(m)$ is a polynomial  of degree $n$ in $m$, and that,
if this property holds for all $n$, it would explain why the truncated asymptotic expansion of
$\langle S^m_N \rangle $  {\em worsens} for increasing values of $m$.

From Eq.\   (\ref{Snt}) one  finds
\begin{equation}
\label{pseudovar}
\sigma_N(m)
\equiv \langle S^m_N\rangle - \langle S_N\rangle ^m
=
\frac{m}{48}(m-1)\pi^2 (4Dt)^{m/2}  (\ln N)^{m/2-2}
\left[ 1+ {\cal O} \left(\frac{\ln^3\ln N}{\ln N}\right)\right]
\end{equation}
i.e.,
\begin{equation}
\label{pseudoratio}
\frac{\sigma_N(m)}{\langle S_N\rangle ^m}
=
\frac{m(m-1)\pi^2}{48 \ln^{2}N} 
\left[ 1+ {\cal O} \left(\frac{\ln^3\ln N}{\ln N}\right)\right].
\end{equation}

Figure \ref{fig:varratio} is a plot of the numerical and analytical results for
$R(m)\equiv\left[\langle S_N\rangle^m/\sigma_N(m)\right]^{1/2}$ versus $\ln N$.
It is clear that the numerical results closely follow the theoretical
prediction given by  Eq.\  (\ref{pseudoratio}) although some differences are noticeable.
We attribute the difference
to the existence of corrective terms of order $ \ln^3\ln N/\ln N$.

\begin{figure}
\begin{center}
\leavevmode
\epsfxsize = 6.5cm
\epsffile{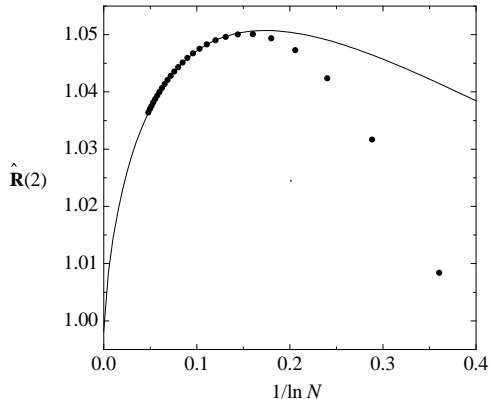}
\end{center}
\caption{
The ratio
$\hat R (2) \equiv  (\protect{\sqrt{24}}/\pi)\ln (N) \,
\sigma_{N}/\langle S_N\rangle  $
versus $1/\ln N$.
The symbols correspond to the numerical evaluation for $N=2^4,2^5,\ldots,2^{30}$.
The line is a curve of the form $a(1+\ln^{-1}N\sum_{j=0}^3 b_j\ln^j\ln N)$ fitted to the last fifteen simulation
 points, i.e., the points corresponding to $N=2^{16}, 2^{17},\ldots,2^{30}$. The fitting parameters are
  $a=0.998$, $b_0=0.009$, $b_1=-0.019$, $b_2= 0.094$ and $b_3=0.005$.
}
\label{fig:vardetalle}
\end{figure}

Let us look
at this question more closely.  In Fig.\  \ref{fig:vardetalle} we have plotted
the ratio
\begin{equation} \label{coefvar}
\hat R(2)\equiv\frac{\sqrt{24}}{\pi} \ln(N) \frac{\sigma_N}{\langle S_N\rangle}
\approx
1+ {\cal O} \left(\frac{\ln^3\ln N}{\ln N}\right)
\end{equation} 
evaluated numerically versus $1/\ln N$, where $\sigma^2_N\equiv\sigma_N(2)$ is the variance.
We note that, at first sight, there are two features that cause unease in this figure: first, the numerical results
are relatively far from unity (the main asymptotic term) even for very large values of $N$, and second,
it is difficult to state that unity is the final value for $N\rightarrow \infty$ by only looking at
the points (the numerical results).
We can shed some light on these two aspects by considering the form and value of the  corrective terms of Eq.\   (\ref{coefvar}).
The first aspect can be understood by taking into account that the functions $\ln^n(\ln N)/\ln N$
have non-negligible values and are only (slowly) decreasing functions for very large values
of $N$ (e.g., when $n=3$, this function is only decreasing for $N\gtrsim 5\times 10^8$).
In order to resolve the second difficulty, we use our knowledge of the form of the first corrective
term of Eq.\   (\ref{coefvar}),  namely, $\sum_{j=0}^3
a_j^{(3)} \ln^j\ln (N) /\ln N$, to  fit the numerical results of the quantity
$\sqrt{24} (\ln N) \sigma_N/(\pi \langle S_N\rangle)$  to the expression $a(1+\ln^{-1}N\sum_{j=0}^3 b_j\ln^j\ln N)$.
Neglecting, for example, the values corresponding to $N=2^0,\ldots,2^{10}$
(recall that our expressions are asymptotic expressions valid for large $N$), the fitted function leads
to $a\simeq 0.988$, a result that is in excellent agreement with our predicted value of unity [i.e., the main term of Eq.\   (\ref{coefvar})].

Finally, it is interesting to note that Eq.\   (\ref{pseudovar}) tells us that, up to {\em first order} in $1/\ln N$,
we can get the $m$-th moment $\langle S^m_N\rangle $  from only the knowledge of the
 first moment $\langle S_N\rangle$.
To be more precise, let us define the parameters $\hat s_j^{(n)}(m)$ through the relationship
\begin{equation}
\langle S_N \rangle^m \approx
\left[4Dt \ln(N) \right]^{m/2} [1-\widehat\Delta(m)]
\label{hatSnt}
\end{equation}
where 
\begin{equation}
\widehat\Delta(m)=\sum_{n=1}^\infty \ln^{-n}N\sum_{j=0}^n 
\hat s_j^{(n)}(m) \ln^j\ln N .
\label{hatDelta}
\end{equation}
Hence,  using Eqs.\   (\ref{Snt})-(\ref{s10}), we find (i) that $s_j^{(n)}(m)=\hat
s_j^{(n)}(m)$ for $j=0,1$ when $n=1$ and for $j=1,2$ when $n=2$, and (ii) that
the first terms of $\langle S_N^m\rangle $ and $\langle S_N\rangle^m $ that are
different are {\em second-order} terms, namely, those corresponding to $j=0$ and
$n=2$:  $\hat s_0^{(2)}(m)-s_0^{(2)}(m)=m(m-1)\pi^2/48$.

It is also worthwhile noting that, working up to second-order asymptotic
corrective terms ($n=2$), the $j$-th  cumulant with $j>2$ is zero, or, in other
words, that the distribution of $S_N$ is, up to this second asymptotic order,
Gaussian.

\section{One-sided trapping problem with a set of random walkers}
\label{NRW}

From Eq.\ (\ref{PhiNInt}) and after integrating by parts,  we can write the survival probability
of a set of $N$ independent random walkers in the one-dimensional one-sided
problem as
\begin{equation}
\label{PhiNx}
\Phi_N(t)= x \int_0^\infty e^{-x \xi} \left[ \text{erf}\left(
\xi \right)
\right]^N d \xi
\end{equation}
where $x=\sqrt{4 D t} \ln (1/p)$ and $\xi=r/\sqrt{4 D t}$.

The asymptotic behaviour of $\Phi_N(t)$ for an arbitrary number of particles
and $x \rightarrow \infty$ is a direct consequence of Watson's lemma \cite{Wong} and
the expansion of the error function $\text{erf}(\xi)$ for small $\xi$:
\begin{equation}
\label{PhiNas}
\Phi_N(t)=\left( \frac{2}{\sqrt{\pi}} \right)^N \left[ \frac{N!}{x^N}-
\frac{(N+2)! N}{3 x^{N+2}}+\ldots\right]
\end{equation}
For $N=1$ we recover the result in Eq.\   (\ref{PhiN1as}).
An slow power-law time decay of $\Phi_N(t)$ is observed in this limit, $\Phi_N(t) \approx 2^N N!/(\sqrt{4 \pi D}
\ln (1/p))^N t^{-N/2}$.

It is interesting to note that the present one-sided multiparticle Rosenstock
trapping problem is related to one of the predator-prey
problems discussed by  Krapivsky and Redner \cite{KR,KR-Am} in which a static
prey or ``lamb'' is captured by one of a set of $N$ diffusing predators or
``pride of lions''. These authors considered the case of $N$
predators and a prey at given relative positions so that the case we study here
differs in the sense that the traps (or preys) are randomly distributed. In
their analysis, Krapivsky and Redner found that the survival probability of the
``lamb'' is given by the power law $t^{-N/2}$.
So, this behaviour agrees with  that of our stochastic ``lamb'' problem in the
long-time regime, $x\gg1$, i.e.,   $t \gg \ln N/[\ln(1-c)]^2$.
The reason for this behaviour is that, for very long times,
how the traps are configured is not essential in the trapping kinetics, and the
the slow $\Phi_N(t) \sim t^{-N/2}$ power law  decay of the fixed  tramp (or ``lamb'') case settles down.

Next,  we will deal with the asymptotic behaviour of $\Phi_N(t)$ for small $x$ and large $N$.
In order to implement the Rosenstock approximation, we will use the expressions for the average one-sided
span  $\langle S_N(t) \rangle$ and its variance  $\sigma_N(2)=\sigma_N^2$
given by Eqs.\   (\ref{Snt1}) and (\ref{pseudovar}).
Then, the zeroth-order Rosenstock approximation given by Eq.\   (\ref{Ros:zero}) is
\begin{equation}
\label{Ros:app0}
\begin{array}{rcl}
\ln\Phi_N^{(0)}(t) &=& \left( 4 D t \ln N \right)^{1/2} \ln(p)
\left[1-\displaystyle\frac{\ln \ln N-2 \omega}{4 \ln N} \right. \\
\noalign{\smallskip}
& & \left. - \displaystyle \frac{\case{1}{4}\displaystyle\ln^2 \ln N
-\left( 1+\omega \right)  \ln \ln N+\pi^2/6+\omega^2+2\omega+2}{8 \ln^2 N}+
{\cal  O}\left(\displaystyle\frac{\ln^3 \ln N}{\ln^3 N}\right)\right]  .
 \end{array}
\end{equation}
In section \ref{OneSidedRosensock} it was mentioned that the relative error for
$\Phi_N^{(0)} $ is of order ${\cal  O}\left(\sigma_N^2 \ln^2p\right)$, so that  the condition
$\sigma_N^2 \ll 1/(\ln p)^2$ must be fulfilled for  the zeroth-order Rosenstock
approximation to be reasonable.
Using the expression for $\sigma_N^2$  given
by Eq.\   (\ref{pseudovar}),  one sees that the relative error goes as
$  \pi^2 D t/[6(\ln p)^2 \ln N]$, so that  $\Phi_N^{(0)}(t) $ should give good results when
$t \ll \ln N/(\ln p)^2$ or $x \ll \sqrt{\ln N}$. Consequently, there exists a time
regime, which becomes {\em larger} as the number of random walkers {\em increases} and the concentration
of traps decreases, where the approximation for the survival probability in
Eq.\   (\ref{Ros:app0}) must work reasonably well.
We will denote by
$\Phi_N^{(0m)}(t)$, $m=0,1,\ldots$ , the zeroth-order Rosenstock approximation for the
survival probability that results from retaining $m$ corrective terms in the
expression for the average one-sided span and ignoring the rest of the terms.
For example,
\begin{equation}
\Phi_N^{(01)}(t) =
\exp\left[-\displaystyle x \left(\ln N\right)^{1/2} \left(1-
\displaystyle\frac{\ln \ln N-2\omega}{4 \ln
N}\right)\right]\; .
\end{equation}
Keeping all the explicit terms in Eq.\  (\ref{Ros:app0}), we get $\Phi_N^{(02)}(t)$.

Proceeding as above, the approximation given by Eq.\   (\ref{Ros:first}) that
includes the variance term, i.e., the first-order  Rosenstock approximation, is
\begin{equation} \label{Ros:app1}
\begin{array}{rcl}
\ln\Phi_N^{(1)}(t) &=& \left( 4 D t \ln N \right)^{1/2} \ln (p)
\left[1-
\displaystyle\frac{\ln \ln N-2 \omega}{4 \ln N} \right. \\
\noalign{\smallskip}
& & \left. - \displaystyle\frac{\displaystyle \case{1}{4} \ln^2 \ln N-
\left( 1+\omega \right) \ln \ln N+\pi^2/6+\omega^2+2\omega+2}{8 \ln^2
N}+{\cal O}\left(\displaystyle \frac{\ln^3 \ln N}{\ln^3 N}\right)\right] \\
\noalign{\smallskip}
& &+\left( 4 D t \ln N \right) \ln^2 (p)
\left[\displaystyle\frac{\pi^2}{48 \ln^2 N}+{\cal O}\left(
\displaystyle\frac{\ln^3 \ln N}{\ln^3 N} \right) \right] .
\end{array}
\end{equation}
We have seen in Sec.\  \ref{OneSidedRosensock} that one can estimate the error of
the first-order Rosenstock approximation by looking at the value of $\kappa_3 (\ln
p)^3 $. The kurtosis, $\kappa_3$, was not calculated in section
\ref{onesidedN} explicitly, but we know (see the last paragraph of that
section) that it is zero up to at least second-order corrective terms, i.e.,
at least, $\kappa_3={\cal O}\left(t^{3/2} \ln^{3}\ln N
\ln^{-3/2}N\right)$.
This means that the relative error of the first-order generalized Rosenstock approximation is a
quantity of order ${\cal O}[(x/\sqrt{\ln N})^3 \ln^{3}\ln N]$ which will be much smaller
than that corresponding to Eq.\  (\ref{Ros:zero}), i.e., ${\cal
O}(x/\sqrt{\ln N})$, as long as the condition $x \ll \sqrt{\ln N}$ is fulfilled.

\begin{figure}
\begin{center}
\leavevmode
\epsfxsize = 6.5cm
\epsffile{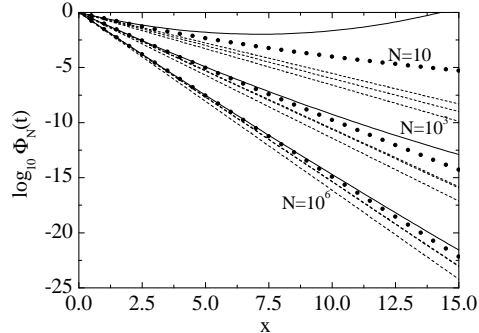}
\end{center}
\caption{Survival probability of $N=10$, $N=10^3$ and $N=10^6$ random
walkers in the one-dimensional one-sided Rosenstock trapping model versus
$x=\protect{\sqrt{4 D t}} \ln (1/p)$.
The circles correspond to a numerical integration
of the exact result, the broken lines correspond to the Rosenstock
approximations of orders $\Phi_N^{(00)}$,  $\Phi_N^{(01)}$,  $\Phi_N^{(02)} $ enumerated from below,
and the solid line corresponds to  $\Phi_N^{(1)}$.  The lines corresponding to  $\Phi_N^{(01)}$ and  $\Phi_N^{(02)} $
for $N=10^6$ are indistinguishable.
 \label{PhiN:RosvsEx}}
\end{figure}

In Fig.\  \ref{PhiN:RosvsEx} we
compare the different order Rosenstock approximations with the numerical evaluation of 
Eq.\   (\ref{PhiNx}) for $N=10$, $10^3$ and $10^6$.
We observe that adding more corrective terms increases the agreement of the Rosenstock approximation
with the numerical result for small values of $x$.   Quite noticeably,  the approximations also get better as
$N$ increases.

\section{Lifetime in the one-sided trapping problem}
\label{sec_lifetime}

Let  the lifetime of the set of $N$ independent random walkers  $T_N$ be defined as
the time at which some random walker of this set is first trapped  or,
conversely, the time at which the lamb is killed by the pride of $N$ lions if
the expression coined by Krapivsky and Redner is used \cite{KR,KR-Am}).
The $m$-th moment of the lifetime distribution is given by
\begin{equation}
\label{avTNm}
\left\langle T_N^m \right\rangle = - \int_0^{\infty} t^m \frac{d \Phi_N (t)}{d t} d t\; .
\end{equation}
or, taking into account Eq.\  (\ref{PhiNx}),
\begin{equation}
\label{avTNmii}
\left\langle T_N^m \right\rangle=\ln (1/p) \int_0^{\infty} d r p^r \overline{t_N^m}(r)
\end{equation}
where $\overline{t_N^m}(r)$ is the $m$-th moment of the time to first reach the
distance $r$ by the first random walker of a set $N$ independent diffusing
random walkers:
\begin{equation} \label{bartNm}
\overline{t_N^m}(r) = -\int_0^{\infty} t^m \frac{d}{d t} \left[ \text{erf}
\left( \frac{r}{\sqrt{4 D t}} \right) \right]^N d t
\end{equation}
Of course, $\overline{t_N^m}(r)$ can be understood as the moments of the
lifetime in the trapping problem with  a fixed trap at a distance $r$ from the
starting site of $N$ independent random walkers. This problem has been widely
studied. For example, Lindenderg {\em et al.} \cite{JSP-Lindenberg} studied the
first passage time for small $N$, finding that the $m$-th moment of
the first-passage-time distribution for the first of the walkers to reach $r$ is finite
if there are at least $2 m + 1$ random walkers starting from the origin.
Hence, by using Eq.\   (\ref{avTNmii}), we can extrapolate these conclusions to the
one-dimensional one-sided Rosenstock's trapping problem. In particular, this means
that $\langle T_1\rangle$ and $\langle T_2\rangle$ are infinite but $\langle
T_N\rangle= C_N/(2D\ln^2p)$ is finite for every $N\geq 3$, the coefficient
$C_N$  being given by $\int_0^\infty \left[\text{erf}(t^{-1/2})\right]^N dt$.

The trapping problem with  fixed trap and large $N$ has also been studied \cite{JSP-Weiss,JSPPRLPRE-Yuste}.
In particular, for the one-dimensional lattice  \cite{JSPPRLPRE-Yuste}
\begin{equation}
\label{YustetNm}
\overline{t_N^m}(r)=\left( \frac{r^2}{4 D} \right)^m \frac{1}{\ln^m \lambda_0 N}
\left\{ 1 + \frac{m}{\ln \lambda_0 N} \left( \frac{1}{2} \ln \ln \lambda_0 N
-\gamma \right)+\ldots \right\}
\end{equation}
where $\lambda_0=1/\sqrt{\pi}$. Then, from Eqs.\   (\ref{avTNmii}) and
(\ref{YustetNm}) we have
\begin{equation}
\label{TNm}
\left\langle T_N^m \right\rangle = \overline{t_N^m}\left( \frac{\left[(2 m)!\right]^{1/2 m}}{\ln 1/p}
\right)\; .
\end{equation}
Thus, the $m$-th moment of the lifetime $\langle T^m_N \rangle$
coincides with the $m$-th moment of the
first passage time to a fixed trap at a distance $\left[ \left( 2 m\right) !
\right]^{1/2 m}$ times the average distance $\langle l \rangle=-1/\ln p$ of a trap to the origin in the
one-sided Rosenstock model with trap density $c=1-p$.
According to Eqs.\   (\ref{TNm}) and (\ref{YustetNm}), and
writing the expansion in terms of $\ln N$ instead of $\ln \lambda_0 N$, we
finally have
\begin{equation}
\label{TNmexp}
\begin{array}{rcl}
\left\langle T_N^m \right\rangle &=&
 \displaystyle\frac{(2 m)!}{[4 D (\ln p )^{2}\ln N ]^m}
 \left\{1+
\displaystyle\frac{m}{\ln N} \left(\displaystyle\frac{1}{2} \ln \ln N-\omega \right)
+\displaystyle\frac{m}{2\ln^2 N} 
    \left[1+\displaystyle\frac{(m+1) \pi^2}{6}  \right. \right. \\
\noalign{\smallskip}
& &\left. \left. +(m+1)\omega^2 + \omega
-\left(m \omega+\omega +\displaystyle\frac{1}{2} \right) \ln \ln N+\displaystyle\frac{(m+1)}{4}
\ln^2 \ln N \right]+{\cal O}\left(\displaystyle\frac{\ln^3 \ln N}{\ln^3 N} \right) \right\}       .
\end{array}
\end{equation}
The asymptotic
expansion for the variance of the distribution of first passage times, $\sigma_N^2=\left\langle
T_N^2 \right\rangle-\left\langle T_N \right\rangle^2$, is calculated from
Eq.\   (\ref{TNmexp}):
\begin{equation}
\label{sigmaN}
\begin{array}{rcl}
\sigma_N &=& \displaystyle\frac{\sqrt{20}}{4 D \left( \ln p \right)^2}
\displaystyle\frac{1}{\ln N} \left\{1+\displaystyle\frac{1}{\ln N} \left(
\displaystyle\frac{1}{2} \ln \ln N-\omega \right)+\displaystyle\frac{1}{2 \ln^2 N}
\left[1+\displaystyle\frac{8 \pi^2}{15} \right. \right. \\
\noalign{\smallskip}
 & & \left. \left. +\omega (1+2 \omega)-\left(\displaystyle\frac{1}{2}
+2 \omega \right) \ln \ln N+\displaystyle\frac{1}{2} \ln^2 \ln N
\right]+{\cal O}\left(\displaystyle\frac{\ln^3 \ln N}{\ln^3 N} \right) \right\} \; .
\end{array}
\end{equation}
The large size of the variance is remarkable: notice that the coefficient of variation
$
\langle T_N \rangle/\sigma_N\approx 1/\sqrt{5}$  (for $N\gg 1$)  is less than unity.

\begin{figure}
\begin{center}
\leavevmode
\epsfxsize = 6.5cm
\epsffile{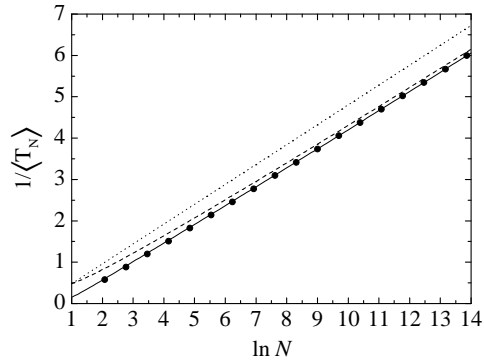}
\end{center}
\caption{Inverse of the lifetime of a set of $N$ independent random walkers
in the one-dimensional one-sided Rosenstock trapping model with a density of
traps $c=1/2$ versus $\ln N$. Numerical integration results are plotted as circles
and the  lines are theoretical asymptotic predictions of zeroth (dotted), first
(broken), and second orders (solid).  The value $D=1/2$ has been used.
\label{InvTN}}
 \end{figure}

 \begin{figure}
\begin{center}
\leavevmode
\epsfxsize = 6.5cm
\epsffile{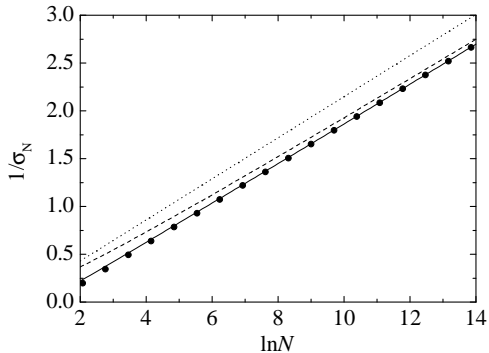}
\end{center}
\caption{The same as Fig.\  \protect\ref{InvTN} but for the inverse of the
square-root of the variance, $1/\sigma_N$.
\label{InvSigmaN}}
\end{figure}

In Figs.\  \ref{InvTN} and \ref{InvSigmaN} we have plotted the inverse
of the average first passage time, $1/\langle T_N \rangle$, and the inverse
of the square-root of the variance, $1/\sigma_N$, respectively, versus $\ln N$.
The second-order approximations for the first passage time in Eq.\  (\ref{TNmexp}) and the
variance in Eq.\  (\ref{sigmaN}) are in excellent agreement with the results of
the numerical integration of Eq.\  (\ref{avTNmii}).

\section{Conclusions and remarks}
\label{conclusions}
In this paper we have studied the one-dimensional one-sided
multiparticle Rosenstock trapping model in which  $N$ independent
random walkers start their random exploration at the same site, $x=0$, of a
one-dimensional lattice whose right side, $x > 0$, has been randomly filled
with static traps with a density $c \equiv 1 - p$.
This problem can be seen either as a multiparticle version of the usual ``two-sided''
trapping  model, or as ``random'' version of the ``predator-prey'' problems
such as that discussed in Refs.\ \cite{KR,KR-Am} regarding the survival
probability of the prey, or in Ref.\  \cite{JSP-Lindenberg} regarding
the prey's lifetime.
Our main interest has been the calculation of the survival probability $\Phi_N(t)$
and the moments of the lifetime $\langle T_N^m \rangle$.

For evaluating $\Phi_N(t)$ for large  $N$ we  resorted to the
(extended) Rosenstock approximation. This required, as an intermediate step but
interesting in itself, the asymptotic evaluation of the moments of the
one-sided span $S_N(t)$  in terms of a series of the form
$\langle S^m_N \rangle
 \approx
 \left(4Dt \ln N \right)^{m/2} \left[1-\sum_{n=1}^\infty
(\ln N)^{-n} \sum_{j=0}^n s_j^{(n)} \ln^j\ln N \right].
$
The importance of evaluating the corrective terms  even for very large values of $N$
is evident as these terms decay mildly as powers of $1/\ln N$.
It is worth mentioning that we have  found  that
 $\langle S^m_N(t) \rangle=\langle S_N(t) \rangle^m$ up to {\em first-order} corrective terms ($n=1$).
The agreement of the Rosenstock approximation with  numerical results for $\Phi_N(t)$
 improves when $N$ increases  and  $x=-\sqrt{4 D t} \ln (1-c)$ decreases,  the agreement being very good when
 $x\ll \sqrt{\ln N}$.
 However,  the stretched exponential behaviour of
$\Phi_N(t)$ characteristic of the Rosenstock approximation breaks down for $x
\gg \sqrt{\ln N} $ and an algebraic long-time tail behaviour settles in:
$\Phi_N(t) \sim t^{-N/2}$.
This is the fluctuation slowdown reported in the trapping problem
literature \cite{DV} and known as the Donsker-Varadhan limit, but here more dramatic,
being algebraic instead of stretched exponential, because of the infinite
half-line free of traps that exists  in the one-dimensional one-sided trapping model.

Once  the survival probability was determined, we
dealt with the problem of the first passage time $T_N$ of the first
random walker of a set of $N$ to the nearest site occupied by a trap. We 
found that the $m$-th moment  $\langle T_N^m \rangle$ coincides with the  $m$-th moment  of the
first passage time to a single fixed trap  placed at the distance $r_m=[(2 m)!]^{1/(2 m)} \langle l \rangle$ from
the origin,  $\langle l \rangle=1/\ln (1/p)$  being the
average distance to the origin of the nearest trap of the random distribution.

Finally, it must be remarked that the extended
Rosenstock approximation could also be used for the multiparticle trapping problem on the
``two-sided'' one-dimensional lattice and on other Euclidean and fractal substrates because
asymptotic series for the average number of
distinct sites visited  $\langle \widetilde S_N(t)\rangle$ have previously been derived for these media
\cite{PRE-RCEuFrac}.
However, the calculation of higher order moments of $\widetilde S_N(t)$
poses a problem of completely different order of magnitude that still remains 
unsolved.  On the other hand, we have found that, up to first-order corrective terms,
 $\langle S^2_N(t) \rangle=\langle S_N(t) \rangle^2$.
At this point, it is tempting to conjecture that this holds  for   $\widetilde S_N(t)$ too.
Were this true, the variance of    $\widetilde S_N(t)$ would be a quantity of {\em second} order and,
therefore, the zeroth-order
Rosenstock approximation with $\langle \widetilde S_N\rangle $ given up to first-order corrective term should lead to a good approximation
for the survival probability $\Phi_N(t)$  for  Euclidean and fractal media too.
Work is in progress to check
these conjectures and thereby extend the results of this present work to more
general trapping problems.

\acknowledgments
This work has been supported in part by the Ministerio de Ciencia y Tecnolog\'{\i}a  (Spain)
through Grant No. BFM2001-0718.
SBY is also grateful to the DGES (Spain) for a sabbatical grant (No.\ PR2000-0116)
and to  Prof.\ K.\ Lindenberg and the Department of Chemistry of the University of California San Diego
for their  hospitality.




\begin{references}
\bibitem{Hughes}  B. H. Hughes, Random Walks and Random Environments,
Volume 1: Random Walks, Clarendon  Press,  Oxford, 1995.

\bibitem{Weiss} G. H.  Weiss,  Aspects and
Applications of the Random Walk,  North-Holland, Amsterdam, 1994.

\bibitem{HollanderWeiss} F. den Hollander and G. H. Weiss, Contemporary
Problems in Statistical Physics, ed. G. H. Weiss, SIAM, 1994, Philadelphia.

\bibitem{ShDba} S. Havlin  and D. Ben-Avraham, Adv. Phys.  36 (1987)  695.

\bibitem{BeelerDamask} J. R.  Beeler, Phys.  Rev.  134 (1964)  1396.
 P. Damask  and  P. Dienes, Point Defects in Metals, Gordon and Breach, New York, 1964.

\bibitem{Rosens} H. B. Rosenstock,  Phys. Rev.  187 (1969) 1166.

\bibitem{Montroll} E. W. Montroll, J. Phys.  Soc.  Jap.  Suppl.  26 (1969)  6.  E. W.  Montroll, J. Math. Phys.  10 (1969) 753.

\bibitem{Oshanin}  G. Oshanin, S.  Nechaev, A. M.  Cazabatand and M. Moreau, Phys. Rev. E   58 (1998)  6134.
S.  Nechaev, G.  Oshanin   and A. Blumen, J. Stat. Phys.  98 (2000)  281.

 \bibitem{Miyagawa}  H. Miyagawa, Y.  Hiwatari, B.  Bernu and  J. P. Hansen, J. Chem. Phys. 88 (1988)  3879.
 T. Odagaki, J. Matsui and Y.  Hiwatari, Phys. Rev. E 49 (1994)  3150.

\bibitem{Klafter}  J. Klafter, A. Blumen and G.  Zumofen, J. Stat. Phys.  36 (1984) 561,  and references therein.

\bibitem{Zumofen} G.  Zumofen and A. Blumen,  Chem. Phys. Lett. 83  (1981)  372.

\bibitem{Blumen} A. Blumen,  J. Klafter  and G. Zumofen, Phys. Rev. B  28 (1983)  6112.

\bibitem{SA} G. M. Sastry and N. Agmon, J. Chem. Phys.  104 (1996) 3022.

\bibitem{KR}  P. L. Krapivsky and  S.  Redner, J. Phys. A: Math. Gen.  29 (1996) 5347.

\bibitem{KR-Am}   P. L. Krapivsky and  S.  Redner,  Am. J. Phys. 67 (1999) 1277.

\bibitem{LarraldeNydemas}
H. Larralde, P.  Trunfio, S. Havlin, H. E. Stanley and G. H. Weiss, Nature (London)  355 (1992) 423.
 G. H. Weiss, I. Dayan, S. Havlin, J. E. Kiefer, H. Larralde,  H. E. Stanley and P. Trunfio, Physica A 191 (1992) 479.
A. M. Berezhkovskii, J. Stat. Phys. 76 (1994) 1089

\bibitem{LarraldeP}
H. Larralde, P.  Trunfio, S. Havlin, H. E. Stanley and G. H. Weiss, Phys. Rev. A 45 (1992) 7128.

\bibitem{PRE-RCEuFrac}
S. B. Yuste and L. Acedo, Phys. Rev. E  60 (1999) R3459.
S. B. Yuste and L. Acedo, Phys. Rev. E  61 (2000) 2340.
S. B. Yuste and L. Acedo, Phys. Rev. E  63 (2001) 011105.

\bibitem{Abramo} M. Abramowitz and I. Stegun I (ed.), Handbook of   Mathematical Functions, Dover, New York, 1972.

\bibitem{Wong} R. Wong, Asymptotic Approximations of Integrals, Academic Press, San Diego, 1989.

\bibitem{JSP-Lindenberg} K. Lindenberg, V. Seshadri, K. E. Shuler and G. H. Weiss, J. Stat. Phys.  23 (1980)  11.

\bibitem{JSP-Weiss} G. H. Weiss, K. E. Shuler and K. Lindenberg, J. Stat. Phys. 31 (1983) 255.

\bibitem{JSPPRLPRE-Yuste} S. B. Yuste and K. Lindenberg,  J. Stat. Phys.  85 (1996) 501.
 S. B.  Yuste, Phys. Rev. Lett. 79 (1997) 3565.
 S. B. Yuste, Phys. Rev. E  57 (1998)  6327.

\bibitem{DV}  D. V. Donsker and S. R. S. Varadhan, Commun. Pure Appl. Math.  28 (1975) 525.
A. A. Ovchinnikov and  Y. B. Zeldovich, Chem. Phys.  28 (1978)  215.
P. Grassberger and I. Procaccia, J. Chem. Phys. 77 (1982)  6281.
R. F. Kayser and J. B.  Hubbard, Phys. Rev. Lett. 51 (1983) 79.

\end{references}
\end{document}